\documentclass[aps,prl,twocolumn,superscriptaddress,floatfix,amsmath]{revtex4}

\usepackage{graphicx}

\begin{document}

\title{Spin-wave propagation in a microstructured magnonic crystal}

\author{A. V. Chumak}

\email{chumak@physik.uni-kl.de}

\author{P. Pirro}

\author{A. A. Serga}

\affiliation{Fachbereich Physik, Nano+Bio Center, and Forschungszentrum OPTIMAS, Technische Universit\"at
Kaiserslautern, 67663 Kaiserslautern, Germany}

\author{M. P. Kostylev}

\author{R. L. Stamps}

\affiliation{School of Physics, University of Western Australia, Crawley, Western Australia 6009, Australia}

\author{H.~Schultheiss}

\author{K. Vogt}

\author{S.~J. Hermsdoerfer}

\author{B. Laegel}

\author{P. A. Beck}

\author{B. Hillebrands}

\affiliation{Fachbereich Physik, Nano+Bio Center, and Forschungszentrum OPTIMAS, Technische Universit\"at
Kaiserslautern, 67663 Kaiserslautern, Germany}

\date{\today}

\begin{abstract}
Transmission of microwave spin waves through a microstructured magnonic crystal in the form of a
permalloy waveguide of a periodically varying width was studied experimentally and theoretically. The
spin wave characteristics were measured by spatially-resolved Brillouin light scattering microscopy. A
rejection frequency band was clearly observed. The band gap frequency was controlled by the applied
magnetic field. The measured spin-wave intensity as a function of frequency and propagation distance is
in good agreement with a model calculation.
\end{abstract}


\maketitle

\begin{figure}
\includegraphics[width=0.85\columnwidth]{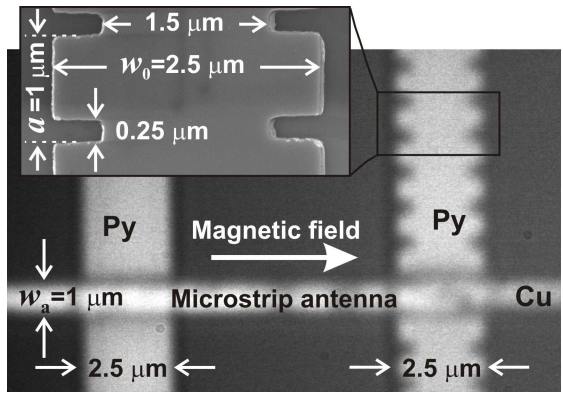}
\caption{\label{fig:epsart1} Scanning electron microscopy and optical images of the structure under
study. The uniform reference waveguide is shown on the left and the magnonic crystal on the right.}
\end{figure}

Magnonic crystals (MCs) operate with spin waves (SW) in the microwave frequency range \cite{Nikitov-2D,
APL-chumak, APL-chumak2, Kalin-soliton, JPD-chumak, Gubbiotti1, Gubbiotti2, Wan09, Kim1, Kim2, Krug,
Krawchyk-3D} and are the magnetic counterpart of photonic and sonic crystals. The greatest success in MC
making has been achieved with yttrium-iron-garnet (YIG) film based structures due to the extremely small
magnetic loss \cite{Nikitov-2D, APL-chumak, APL-chumak2, Kalin-soliton, JPD-chumak}. However,
comparatively large sizes of these devices (hundreds of microns) and the incompatibility of the YIG film
growing process with modern CMOS technology inhibit their wide practical use. Applications in
microelectronics require downscaling to sub-micron sizes and a replacement of YIG films by thin
ferromagnetic metallic layers.

Previous studies of metal film based microstructured magnonic crystals (micro-MCs) were mostly focused
on their thermal SW spectrum \cite{Gubbiotti1, Gubbiotti2, Wan09}, or were purely theoretical
\cite{Kim1, Kim2, Krug, Krawchyk-3D}. Here we report on the experimental observation and
characterization of spin-wave propagation in a metal micro-MC.

To ensure SW propagation the crystal was fabricated as a SW-waveguide made from a permalloy (Py) stripe
with a periodically varying width. This concept was suggested in theoretical works done in the group of
S.~K.~Kim \cite{Kim1, Kim2}. However, the proposed feature sizes of the order of nanometers, which were
chosen due to size limitation in computer simulation, and consequently practically vanishing SW group
velocities make the original structure very difficult to fabricate and test at the current level of
technology. In this work we have modified the original waveguide geometry such that it is now supporting
propagation of spin waves over distances of tens of micrometers. Towards this end we have increased the
waveguide thickness (the SW group velocity is roughly proportional to this dimension) and applied a bias
magnetic field across the waveguide (see Fig.~1) to form conditions for propagation of guided
Damon-Eschbach surface spin waves \cite{waveguide}. In in-plane magnetized metallic samples with a
high-magnetic moment these waves have the highest group velocities. Furthermore, for this magnetization
direction the waveguide's internal magnetic field is strongly inhomogeneous due to static
demagnetization. As a result, in addition to periodical variation of the waveguide width, periodical
modulation of the internal bias field takes place in this geometry. This additionally increases the
efficiency of spin-wave reflection.

Electron beam lithography, molecular beam epitaxy, and lift-off process were used to fabricate the
magnonic crystal in the form of a notched permalloy stripe on thermally oxidized Si(001) substrate (see
Fig.~1). The width of the 40~nm thick Py (Ni$_{81}$Fe$_{19}$) stripe varies periodically between $w_0 =
2.5$~$\mu$m and $w_1 = 1.5$~$\mu$m. The length of the 2.5~$\mu$m-wide sections is 0.75~$\mu$m and the
length of the 1.5~$\mu$m-wide sections (``notches'') is 0.25~$\mu$m. This forms a magnonic crystal with
a lattice constant $a = 1$~$\mu$m. As a reference a second waveguide with a uniform $w_0 = 2.5$~$\mu$m
width was also patterned on the same substrate 6~$\mu$m apart from the magnonic crystal (see Fig.~1).

Spin waves are excited by the microwave Oersted field created by a 500~nm thick and $w_\mathrm{a} =
1$~$\mu$m-wide copper antenna, which is placed across the both Permalloy waveguides (see Fig.~1). In
order to detect spin waves the space-resolved Brillouin light scattering microscopy is used
\cite{mu-BLS}: a focused laser beam probes spin waves with a spatial resolution of 250~nm and a
frequency resolution of 300~MHz \cite{mu-BLS, demidov1, demidov3}.

\begin{figure}
\includegraphics[width=1\columnwidth]{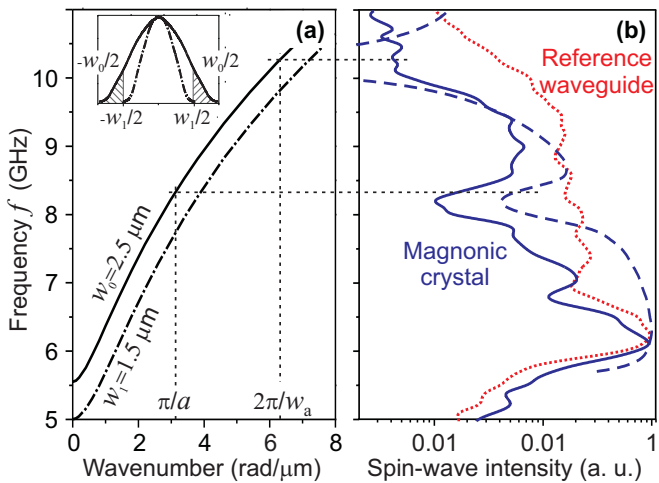}
\caption{\label{fig:epsart2} (Color online).
(a) Calculated SW dispersion curves for uniform waveguides with $w_0 = 2.5$~$\mu$m (solid line) and
1.5~$\mu$m (dash-dotted line). The inset shows profiles of the fundamental width modes in uniform
waveguides of different widths.
(b) Normalized SW intensity measured at the distance $x = 8$~$\mu$m from the antenna versus the applied
frequency. Solid line -- the magnonic crystal; dotted line -- the reference waveguide. Dashed line
represents the calculation for the magnonic crystal. For both panels the bias magnetic field and the
saturation magnetization are 500~Oe and $M_\mathrm{s} = 770$~G.}
\end{figure}

The calculated dispersion for the lowest (fundamental) width mode of a $w_0 = 2.5$~$\mu$m-wide uniform
waveguide is shown in Fig.~2(a) as solid line. For comparison, in this panel we also show the dispersion
for a 1.5~$\mu$m-wide uniform waveguide which corresponds to the narrow sections of the magnonic
crystal. The calculation was carried out by numerically solving the integro-differential equation
derived in Ref.~\cite{waveguide}. Note, that this calculation takes into consideration the inhomogeneity
of the internal static magnetic field and the static magnetization which is also responsible for the
bell-shaped transverse profile of the fundamental mode (see, the inset in Fig.~2).

The experimentally measured SW intensity for the reference waveguide is shown in Fig.~2(b) (dotted line)
as a function of the applied microwave frequency. The maximum intensity corresponds to spin waves with
wavenumbers slightly larger than $k = 0$ because of the highest excitation efficiency and the highest SW
group velocity. With increasing frequency (and increasing spin-wave wavenumber, respectively) the
excitation efficiency drops \cite{demidov3}. For the used antenna it gets close to zero above 11~GHz. A
weak oscillatory intensity variation with frequency can be understood as beating of the fundamental
width mode with the third one, which is also excited by the antenna but less efficiently \cite{demidov1,
serga}.

Solid line in Fig.~2(b) shows the measured SW intensity for the magnonic crystal. A pronounced rejection
band (where spin waves are not allowed to propagate) is clearly observed for frequencies close to 8~GHz.
One can see that the rejection frequency is slightly shifted down with respect to the frequency expected
from the simple Bragg analysis of the SW dispersion for the uniform 2.5~$\mu$m-wide waveguide
($k_\mathrm{rej1} = \pi/a = 3.14$~rad/$\mu$m). We suppose that this is due to inhomogeneity of the
internal magnetic field within the crystal. The decrease in the internal field between the opposite
notches shifts the dispersion curve downwards in frequency. As a result the condition $k_\mathrm{rej1} =
\pi/a$ is fulfilled for a smaller frequency value. It is worth noting that the rigorously calculated SW
intensity, which is also shown in Fig.~2(b), is in good agreement with the experiment. The second
rejection band $k_\mathrm{rej2} = 2\pi/a = 6.28$~rad/$\mu$m is visible in Fig.~2(b) as well. However, it
is not well pronounced because $k_\mathrm{rej2}$ coincides with the edge of the antenna excitation band
$k_\mathrm{max} = 2\pi/w = 6.28$~rad/$\mu$m.

Scattering of spin waves from an array of notches on a stripe waveguide is described by an equation
similar to Eq.~(5) in Ref.~\cite{scattering}:
\begin{eqnarray}
{\hat\chi}(\textbf{r})\textbf{h}_\mathrm{d}(\textbf{r})+\int_{S} \hat{G}_{\mathrm{exc}}(\textbf{r}-\textbf{r}')
\hat{\nu}(\textbf{r}')\textbf{h}_\mathrm{d}(\textbf{r}') d^2\textbf{r}'=\textbf{m}_0(\textbf{r}),
\end{eqnarray}
where $\textbf{h}_\mathrm{d}$ is the total dynamic dipole field of the incident and the scattered spin
waves, $\hat{\nu}(\textbf{r})=({\hat\chi}_0(\textbf{r})^{-1}{\hat\chi}(\textbf{r})-\hat{I})$, $\hat\chi$
is the microwave magnetic permeability tensor which is position dependent through the nonuniformity of
the internal static magnetic field and the static magnetization, ${\hat\chi}_0$ is its value for the
respective uniform waveguide of the width $w_0$, $\hat{G}_{\mathrm{exc}}$ is the Green's function of
excitation of waves in this waveguide, $\textbf{m}_0$ is the amplitude of the spin wave with a
wavenumber $k$ incident on the notch array, and $\hat{I}$ is the identity tensor. This equation allows
an analytical solution in the First Born Approximation (FBA) \cite{JPD-chumak, scattering}. From this
analytic solution it can be shown that the depth of the rejection minima is proportional to the total
dipole energy $E$ of the incident spin wave contained in the waveguide cross-section which is cut out by
the notches (shaded area in the inset in Fig.~2). Figuratively, part of the spin-wave energy incident on
a pair of notches is reflected because the SW width profile does not fit into the narrow waveguide
section. The reflection grows faster than linearly with the notch depth, as $E$ grows with the size of
the dashed area in the inset. In particular, one may expect a negligible rejection when the notch depth
is smaller than the size of the demagnetized area at the edge of the uniform waveguide. This assumption
is confirmed by our measurements on a different micro-MC having 250~nm-deep notches.

To get a closer insight into the wave scattering mechanisms we performed a simulation based on a
phenomenological approach from \cite{APL-chumak, JAP-chumak}. For this purpose we consider the micro-MC
as a periodical sequence of sections of regular transmission lines with different propagation constants
(different $k$-values) for the same carrier frequency. The rejection coefficient at the junction of
wider-to-narrower waveguide can then be written as $\Gamma_{0-1} = (k_1-k_0)/(k_1+k_0) + \Gamma'$, where
$k_0$ and $k_1$ are the wavenumbers for $w_0$-wide and $w_1$-wide waveguide sections, respectively
\cite{JAP-chumak}. As it was mentioned above, the spin wave is scattered back not only due to the
difference in $k$, which is accounted by first summand $\Gamma_{0-1}$, but also because its initial
transverse profile does not fit into the width $w_1$. To account for this additional reflection
mechanism we phenomenologically introduce $\Gamma'$. The rejection coefficient for the waveguide
junction narrower-to-wider contains only one term $\Gamma_{1-0} = - (k_1-k_0)/(k_1+k_0)$.

The theoretical dependence of the SW intensity on the applied frequency is shown in Fig.~2(b) with a
dashed line. It was calculated as $(F(k_0)/|T_{11}|)^2$, where $F(k_0) \propto \sin(k_0 \cdot
w_\mathrm{a}/2)/k_0$ is the efficiency of the antenna excitation. $T_{11}$ is the element (1,1) of the
MC transmission matrix (see \cite{APL-chumak, JAP-chumak}) which includes $\Gamma_{1-0}$,
$\Gamma_{0-1}$, and the experimentally measured spatial SW damping corresponding to a Gilbert damping
parameter $\alpha \approx 0.007$ \cite{Patton-FMR}. This best fit is obtained for $\Gamma'$=0.12. It
means that 12 percent of incident beam energy is reflected back due to the geometrical mismatch between
the waveguide sections. We should also emphasize that the reflection caused by the change of the SW
wavenumbers is of the same order.

The experimentally measured transmission characteristic for the micro-MC is shown in Fig.~3(a) for
different bias magnetic fields. Good agrement between the theory and the experiment is seen. The
first rejection band is clearly visible for all the fields higher than 150~Oe. This value
corresponds to the minimum field one has to apply in order to saturate the magnonic crystal.  One
also sees that variation of the applied bias magnetic field in the range from 150~Oe to 700~Oe
makes it possible to control the band gap frequency in the range from 6.5 to 9~GHz.

\begin{figure}
\includegraphics[width=1\columnwidth]{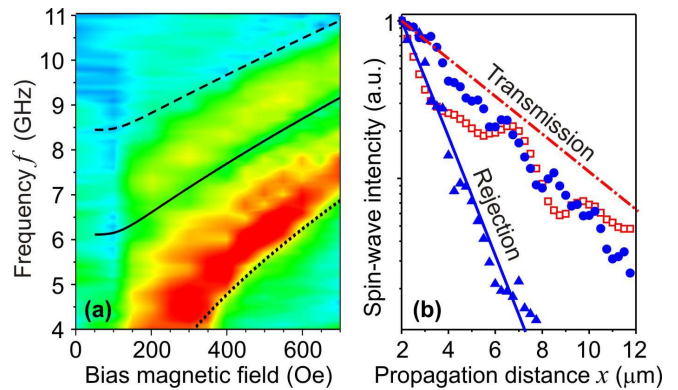}
\caption{\label{fig:epsart3} (Color online).
(a) Measured spin-wave intensity as a function of frequency and bias magnetic field. Logarithmic scale
is used. Red color corresponds to the maximum and blue color to the minimum SW intensity. The dotted
line shows the calculated frequency for the zero SW wavenumber; solid and dashed lines mark the
frequencies for the first rejection band ($k_\mathrm{rej1}$) and for the limit of antenna excitation
($k_\mathrm{max}$), respectively. Distance from the antenna edge $x = 8$~$\mu$m. \\
(b) Measured spin-wave intensity as a function of propagation distance $x$.
Filled triangles -- magnonic crystal rejection band (the applied frequency is 8.1~GHz);
filled circles -- magnonic crystal transmission band (8.9~GHz);
opened squares -- reference waveguide (8.1~GHz).
Solid and dash-dotted lines show the calculation for the magnonic crystal and
the reference waveguide (8.1~GHz).
Bias magnetic field is 500~Oe.}
\end{figure}

Figure~3(b) shows the SW intensity for both the magnonic crystal and the reference waveguide as a
function of the spin-wave propagation distance ($x=0$ corresponds to the edge of the antenna)
\cite{footnote2}. The intensity was measured in the middle of the stripes along their longitudinal axes.
The oscillations of the SW intensities with $x$ can be interpreted as the spatial beating of different
waveguide width modes \cite{demidov1}. The dependence obtained in the transmission band of the micro-MC
is very similar to the one from the reference waveguide. This fact proves the ability of practically
undisturbed SW propagation in the metal waveguide with strongly damaged edges. At the same time spin
waves in the band gap undergo pronounced resonant scattering. It results in an intensity which is ten
times smaller than that for the reference sample after passing eight periods of the structure.

In conclusion, a micro-sized magnonic crystal operational at microwave frequencies has been fabricated
in the form of a notched permalloy waveguide. Formation of pronounced magnonic band gaps was observed.
They are seen as considerable decrease in SW transmission caused by the resonant backscattering from a
periodical lattice. The band gap frequency can be tuned in the range from 6.5 to 9~GHz by varying the
applied magnetic field.

Financial support by the DFG SE 1771/1-1, Australian Research Council, and the University of Western
Australia is acknowledged.


\begin{thebibliography}{sotief}
\bibitem{footnote2} The data for $x<2$~$\mu$m were removed from the picture because magnetization dynamics in this
                    area is predominantly not a propagating wave but represents a near antena field.
\end{thebibliography}
\end{document}